\documentclass[twocolumn,trackchanges]{aastex7}

\usepackage{amsmath}
\begin{document}

\title{The Filament Rift:\\$\Lambda$CDM's Structural Challenge Against Observation}

\author[orcid=0000-0003-0126-8554,sname='Saeed Tavasoli']{Saeed Tavasoli\textsuperscript{*}}
\affiliation{Department of Astronomy and High Energy Physics, Kharazmi University, Tehran, Iran}
\email[show]{\textsuperscript{*}stavasoli@khu.ac.ir}
\author[orcid=0009-0003-2960-1563,sname='Parsa Ghafour']{Parsa Ghafour\textsuperscript{\textdagger}}
\affiliation{Department of Astronomy and High Energy Physics, Kharazmi University, Tehran, Iran}
\email[show]{\textsuperscript{\textdagger}P.Ghafour@outlook.com}

\begin{abstract}

This study presents the first extended comparison of cosmic filaments identified in SDSS DR10 observations ($z < 0.05$) and the IllustrisTNG300-1 $\Lambda$CDM simulation ($z = 0$), utilizing the novel GrAviPaSt filament-finder method. The analyses are performed on both macro- and micro-filaments, each characterized by their length, thickness, and contrast in mass density. In addition to total sample comparisons, two subcategories of micro-filaments, GG (linking galaxy groups) and CC (linking galaxy clusters), are introduced to further analyze discrepancies between the $\Lambda$CDM model and observation. While $\Lambda$CDM produces extended macro-filaments, such structures are largely absent in SDSS, and where present, they exhibit higher densities than their simulated counterparts. Micro-filaments also show notable density discrepancies: at fixed length and thickness, observational filaments are significantly denser than those in the simulation. Employing radial density profiles reveal that micro-filaments in the $\Lambda$CDM simulation exhibit higher contrasts in mass density relative to the background compared to their observational counterparts. Notably, CC type micro-filaments displayed enhanced density contrasts over GG types in the simulation, while observational data showed the opposite trend. Furthermore, SDSS galaxies in both GG and CC micro-filaments exhibit lower specific star formation rates (sSFR) and older stellar populations, while TNG300-1 micro-filaments host more actively star-forming galaxies within the intermediate stellar mass range. These results reveal persistent discrepancies between observational data and the $\Lambda$CDM reconstruction of cosmic filaments, pointing to possible tensions in our current understanding of large-scale structures and their environmental effects on galaxy evolution.

\end{abstract}


\section{Introduction\label{sec:1}}
Observational surveys (e.g., 2dFGRS (\cite{colless20012df}); SDSS (\cite{stoughton2002sloan}); DESI (\cite{adame2024early})), along with insights from numerical simulations (e.g., \cite{springel2005simulations,vogelsberger2014properties}), have firmly established that cosmic matter is distributed non-uniformly, forming an intricate network known as the Cosmic Web (\cite{davis1982survey,bond1996filaments,peebles2020large}). This web comprises the largest nonlinear structures in the Universe, which emerged through gravitational collapse: massive galaxy clusters residing at its nodes and interconnected by filaments, while vast under-dense voids occupy the majority of its volume (\cite{peebles1967gravitational,zeldovich1982giant,cautun2014evolution}).

Across this hierarchical framework, clusters represent the deepest gravitational potential wells, characterized by high galaxy densities, virialized gas, and complex physical processes such as AGN feedback and mergers (\cite{kravtsov2012formation}). Conversely, cosmic voids as the most quiescent regions of the Cosmic Web, offer complementary tests for galaxy formation models in near-isolation (\cite{peebles2001void,tavasoli2015void}).

Cosmic filaments act as crucial environments for testing $\Lambda$CDM cosmology. Their elongated morphology preserves imprints of primordial density fluctuations and influences galaxy evolution through anisotropic matter flows (\cite{cautun2014evolution,zhang2024statistical,wang2024darkai}). These regions play a decisive role in shaping galaxy properties, such as the suppression of star formation in galaxies near filament spines compared to those in voids (\cite{singh2020study}), and the enrichment of circumgalactic media that display kinematic imprints of gas accretion (\cite{vulcani2019gasp,aragon2019galaxy}) as part of a broader pattern of filament-driven galaxy evolution. Statistical characterizations, including filament lengths, density profiles, and magnetic field configurations, could further constrain cosmological models (\cite{malavasi2020characterising,vernstrom2021discovery,galarraga2022relative,galarraga2024evolution}).

While the $\Lambda$CDM model has successfully reproduced many statistical features of the large-scale structure of the Universe, detailed studies of specific components of the Cosmic Web have surfaced several observational challenges. In void environments, the so-called void phenomenon highlights discrepancies between simulations and data, such as the under-representation of low-mass galaxies and the unexpected isolation of massive systems in regions that should be sparsely populated (\cite{peebles2001void,klypin1999missing,tavasoli2013challenge,tavasoli2021void}). In galaxy clusters, significant challenges persist in baryonic modeling. The observed properties of intracluster media, such as the entropy distribution, the AGN feedback efficiency, and the gas dynamics remain only partially understood and deviate from theoretical predictions (\cite{kravtsov2012formation,planelles2014role}). In addition, large-scale velocity flows, measured through galaxy peculiar motions and bulk flow surveys, exhibit amplitudes and coherent patterns that surpass those predicted by linear theory under the $\Lambda$CDM framework (\cite{watkins2009consistently,kashlinsky2008measurement,habibi2018peculiar}). Moreover, a tension has emerged regarding the weighted amplitude of matter fluctuations $S_{8}$ (\cite{hildebrandt2017kids}), with cosmic shear surveys and large-scale structure data exhibiting about $2.5\sigma$ discrepancy relative to the Cosmic Microwave Background anisotropy measurements from Planck (e.g., \cite{aghanim2020planck,he2023s8,sabogal2024quantifying}).
These challenges, embedded within the structure and dynamics of the cosmic web, motivate efforts to more precisely examine the $\Lambda$ CDM cosmology to the observations.

In this study, the novel GrAviPaSt method (\cite{ghafour2025gravipast}) was employed to identify three-dimensional filament catalogs derived from the SDSS DR10 survey (\cite{ahn2014tenth}) and the IllustrisTNG300-1 simulation (\cite{nelson2019illustristng}), as detailed in Sec.~\ref{sec:2}. The structural characteristics of filaments and their influence on the physical properties of galaxies were analyzed to assess the consistency of $\Lambda$CDM predictions with observations. In Sec.~\ref{sec:3}, comparisons were conducted between the lengths and mass density contrasts of cosmic filaments. In addition, both radial and longitudinal density profiles of micro-filaments were examined, revealing potential tensions with theoretical expectations. Key properties of filament galaxies, such as color and star formation rate, were analyzed to probe environmental influences on star formation activity. Notably, two specific filament types bridging dense structures, $GG$ (group-group) and $CC$ (cluster-cluster) filaments, were studied across both SDSS and IllustrisTNG300-1 datasets. Finally, in Sec.~\ref{sec:4}, a summary of this dual observational–simulation framework is presented, highlighting potential discrepancies in large-scale structure formation and providing a direct test of the $\Lambda$CDM model's robustness.

\section{Sample Selection and Methodology\label{sec:2}}
The datasets employed in this study contain carefully curated observational and $\Lambda$CDM simulated samples to investigate the distribution of galaxies within cosmic filaments, enabling a robust comparison between observation and simulation through different filament and filament galaxy characteristics.

\subsection{Observational Sample\label{sec:2.1}}
The observational sample of galaxies is extracted from the SDSS DR10 dataset (\cite{york2000sloan,ahn2014tenth}), which was corrected for motion relative to the Cosmic Microwave Background (CMB) and for peculiar velocity effects by \cite{tempel2014flux}. The sample is volume-limited with an absolute magnitude cut of $M_r \leq -18$ for this study, and is therefore restricted to a redshift range of $z < 0.05$. Stellar masses and spectroscopic properties (due to the implementation of the GrAviPaSt method; see Sec.~\ref{sec:2.3}) are sourced from the MPA-JHU DR7 catalog (\cite{kauffmann2003stellar,brinchmann2004physical,salim2007uv}), and galaxy colors (see Sec.~\ref{sec:3.3}) are obtained from the NYU-VAGC (\cite{blanton2005new}). The final sample includes approximately 50,000 observational galaxies.

\subsection{$\Lambda$CDM Simulation Sample\label{sec:2.2}}
To compare galaxy distributions within cosmic filaments between observation and $\Lambda$CDM cosmology, galaxies from the IllustrisTNG300-1 gravo-magnetohydrodynamical simulation are utilized. This simulation is part of the IllustrisTNG project\footnote{https://www.tng-project.org} (\cite{springel2010pur,nelson2019illustristng}) and represents the highest-resolution run within the IllustrisTNG300 suite. It adopts cosmological parameters informed by Planck cosmology (\cite{ade2016planck}), including density parameters of $\Omega_{m} = 0.3089$ for matter, $\Omega_{b} = 0.0486$ for baryons, and $\Omega_{\Lambda} = 0.6911$ for the cosmological constant, along with $n_{s}$ = 0.9667 for the scalar spectral index, $\sigma_8$ = 0.8159 for the amplitude of fluctuations, and $H_{0} = 100 h$ $(km/s/Mpc)$ with $h_{0}$ = 0.6774 for the Hubble constant. The simulation features dark matter and baryonic mass resolutions of $5.9\times10^7M_\odot$ and $1.1\times10^7M_\odot$, respectively. It encompasses a cubic volume with a side length of approximately 205 $\mathrm{Mpc}/h$ and provides a sufficiently large sample size to enable a statistically robust characterization of filamentary structures. For consistency with the observational sample, galaxies from IllustrisTNG300-1 possessing stellar masses exceeding $10^9 M_\odot$, located at $z = 0$ (snapshot = 99), and exhibiting an absolute magnitude of $M_r \leq -18$ are considered, resulting in a sample of approximately 230000 simulated galaxies.

\subsection{Filament Identification \label{sec:2.3}}
To identify the filamentary structures (micro- and macro-filaments), the GrAviPaSt filament identification method (\cite{ghafour2025gravipast}) is employed. This method leverages gravitational potential, an A*-like path-finder, and the minimum spanning tree (MST) of the central mass distribution of the galaxy systems.

The GrAviPaSt algorithm proceeds through four sequential phases applied to the input distribution. First, it determines the center of mass (COM) of the galaxy systems and uses these COMs to construct a minimum spanning tree (MST) via Prim’s algorithm (\cite{prim1957shortest}), identifying pairs of gravitationally bound systems between which microfilament structures are to be searched (hereafter, microfilaments refer to filaments connecting two such systems). In the second phase, an A*-like path-finding algorithm is employed in combination with the gravitational potential field between each pair of galaxy systems to trace the path of maximum potential depth, which is then considered as the skeleton of the microfilament. The third phase constructs the final microfilament structure based on the spatial distribution of galaxies surrounding the identified skeleton. Finally, the fourth phase groups microfilaments into macrofilaments by chaining them together. This grouping is performed via a three-step procedure that evaluates the thickness, angular alignment, and gravitational potential of microfilaments connected to each node, determining which pairs can be merged into physically consistent macrofilament structures. Thus, microfilaments refer to relatively shorter filamentary structures that directly connect two galaxy systems, acting as gravitational bridges and indicating ongoing or potential matter exchange between the paired systems. In contrast, macrofilaments denote extended filamentary structures that form the backbone of the cosmic web, emerging through the hierarchical assembly of multiple microfilaments. The GrAviPaSt algorithm identifies cosmic filaments across a range of lengths, thicknesses and mass density contrasts ($\delta_{M}$).

\subsection{Cosmic Filament Catalogs \label{sec:2.3.2}}
To identify cosmic filaments within both the SDSS spectroscopic sample and the IllustrisTNG300-1 simulation data, the first step involves identifying bound galaxy systems. For the observational dataset, the friends-of-friends (FOF) group catalog of the sample is employed, as described in \cite{tempel2014flux}. Based on spectroscopic richness, clusters are defined as systems with more than 15 members, while groups are intermediate-density structures containing 4 to 7 members. Galaxies in these environmental subsamples satisfy the same selection criteria as the primary sample ($M_r < -18$ and $z < 0.05$). To ensure comparability, an analogous group and cluster catalog is constructed for the IllustrisTNG300-1 sample using identical richness criteria, enabling the identification of $GG$ and $CC$ micro-filaments in both the observation and simulation samples. Using these group and cluster catalogs, micro-filaments are classified into three categories across the study: $Total$ micro-filaments, containing all micro-filaments in the catalogue; $CC$ micro-filaments, structures connecting two galaxy clusters; and the $GG$ micro-filaments, structures connecting two galaxy groups.

Applying the GrAviPaSt method (\ref{sec:2.3}) yields 273 micro-filaments in the observational sample and 2,786 in the simulation. These structures encompass 796 and 11,994 filament galaxies respectively. Within these catalogs, 11 CC and 88 GG micro-filaments are identified in the observation, compared to 223 CC and 534 GG micro-filaments in the simulation. The CC micro-filaments host 47 galaxies in the observational sample and 1,819 in the simulation, while the GG micro-filaments include 218 and 1,477 galaxies respectively.

\section{Results and Discussion\label{sec:3}}

The GrAviPaSt method (\cite{ghafour2025gravipast}) was applied to the datasets outlined in Section~\ref{sec:2} to investigate the structural and environmental properties of both macro- and micro-filaments, along with their associated filament galaxies. These analyses enabled a direct comparison between observed filamentary characteristics and theoretical predictions derived from the $\Lambda$CDM model.

In the Prim’s MST phase of the GrAviPaSt method, the center of mass coordinates of bound galaxy systems are selected. These systems include groups or clusters with at least four members, reflecting the minimum richness criterion detailed in Section~\ref{sec:2}. These coordinates functioned as MST nodes, with the resulting micro-filament structures forming bridges that connect them across the cosmic web.

\subsection{Macro-Filaments\label{sec:3.1}}

\begin{figure}
	\centering
	\includegraphics[width=0.47\textwidth]{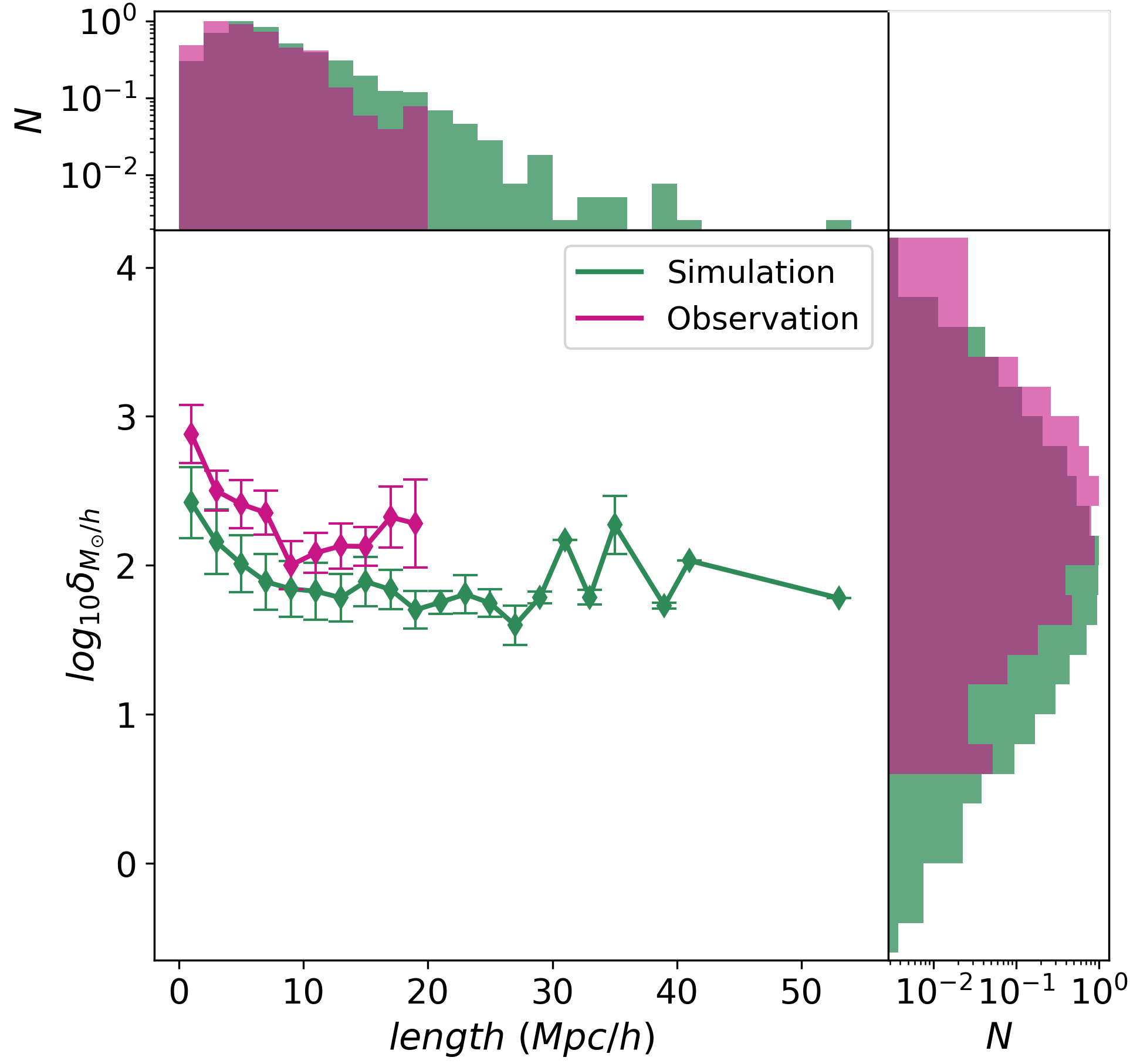}
	\caption{
		Median trends of macro-filament mass density contrast ($\delta_{M_{\odot}/h}$) are shown with respect to filament length for both observational and simulation datasets, with $1\sigma$ scatter represented as error bars. Histograms of the distributions are displayed adjacent to the central plot to highlight the comparative structure of the two samples.
		\label{fig:1}}
\end{figure}

The largest filamentary structures in the cosmic web, known as macro-filaments, are identified by the GrAviPaSt method through the connection of micro-filaments and forming chain-like structures. The filament catalog for the $\Lambda$CDM simulation and the SDSS sample contains 1821 and 217 macro-filaments respectively, each encompassing all filament galaxies included in the catalog, as described in Section~\ref{sec:2.3.2}. These macro-filaments are characterized using two parameters: co-moving length and stellar mass density contrast (hereafter denoted as mass density contrast $\delta_{M_{\odot}/h}$), which is defined as:
\begin{equation}
	\label{eq:1}
	\delta_{m} = \frac{\rho^{Macro}_{m}-\rho^{b}_{m}}{\rho^{b}_{m}}
\end{equation}
where $\rho^{Macro}_{m}$ and $\rho^{b}_{m}$ represent the mass densities of the macro-filament and the background, respectively. The logarithmic background mass densities of the observational data and the $\Lambda$CDM simulation are $8.7$ and $8.6$, respectively; indicating a nearly similar background density of the two samples. These distributions, presented in Fig.~\ref{fig:1}, are compared between observational and simulation datasets.

In Fig.~\ref{fig:1}, markers indicate the median mass density contrast within each co-moving length bin for both datasets. Notably, macro-filaments in the $\Lambda$CDM simulation appear substantially longer and exhibit slightly less pronounced mass density contrasts relative to the background, compared to observational macro-filaments across all length bins.

Although the distribution peaks are closely aligned, observational macro-filaments exhibit a narrower range in both co-moving length and mass density contrast ($\delta_{M_{\odot}/h}$) compared to those derived from the $\Lambda$CDM simulation. Statistical comparisons were performed using two-sided Kolmogorov–Smirnov (K-S) tests. The resulting p-values for the co-moving length and mass density contrast distributions of macro-filaments are $0.02$ and $0.79$, respectively. These values suggest a statistically significant difference in the length distributions, with a slight variation in the mass density contrast.

These discrepancies indicate tension between the macro-filaments in observational data and the reconstruction of these structures by the $\Lambda$CDM model.

\subsection{Micro-Filaments\label{sec:3.2}}

Micro-filaments are defined as matter bridges that connect pairs of bound galaxy systems (i.e., groups and clusters). As mentioned in Sec.~\ref{sec:2.3.2}, in addition to the $Total$ micro-filament catalogs of the observational and $\Lambda$CDM simulation samples, two subcategories are considered in this comparison: $GG$ and $CC$ micro-filaments. Various characteristics of the micro-filaments identified using the GrAviPaSt method (\cite{ghafour2025gravipast}) are examined; including their lengths, thicknesses, and stellar mass density contrast (hereafter denoted as mass density contrast) distributions. Furthermore, radial and longitudinal mass density profiles are probed to provide insights into their structural features.

\begin{figure}
	\centering
	\includegraphics[width=0.47\textwidth]{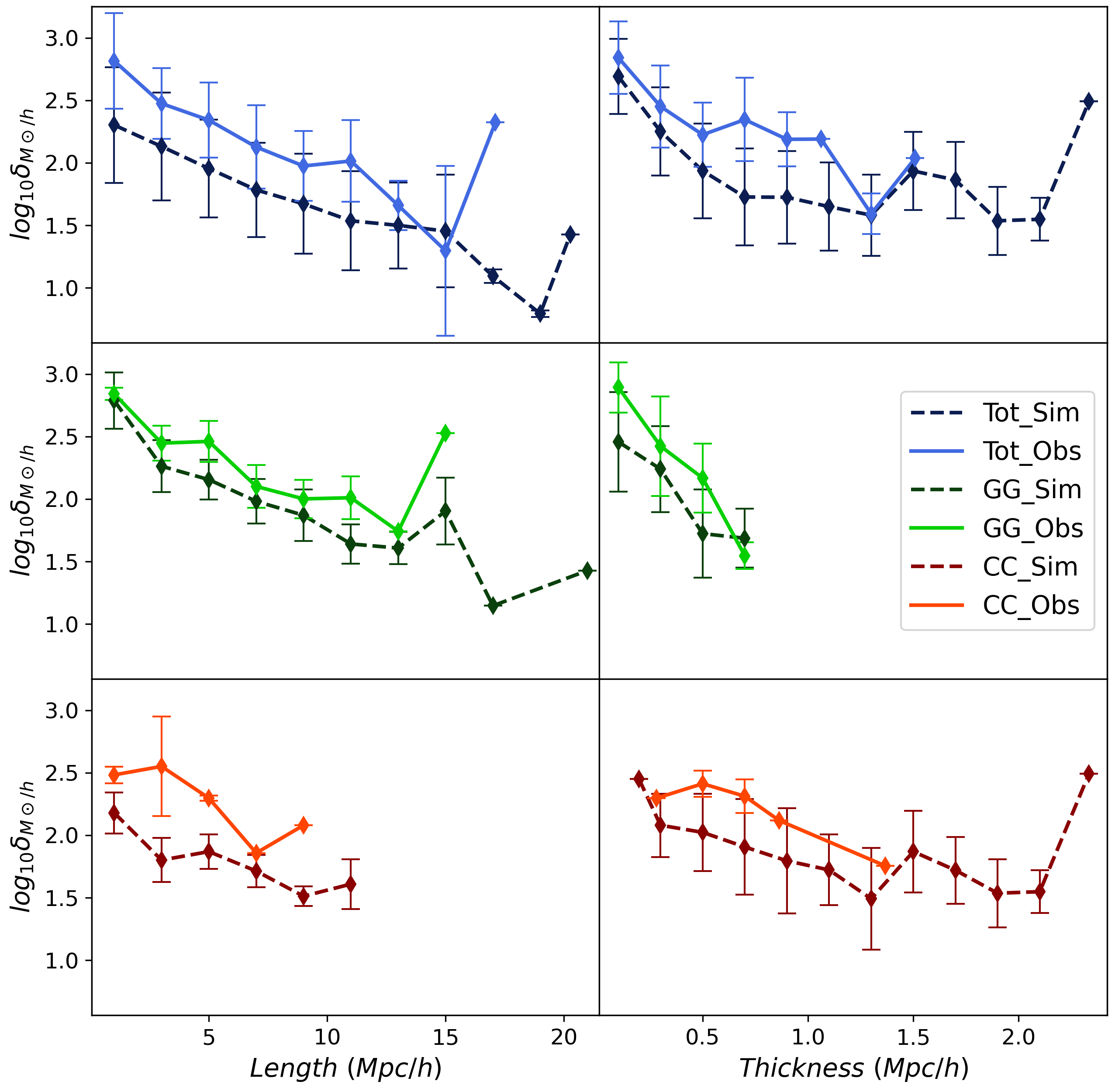}
	\caption{
		Median trends of micro-filament mass density contrast ($\delta_{M_{\odot}/h}$) with respect to length and thickness are shown for the $Total$, $GG$, and $CC$ categories, arranged from top to bottom. Error bars represent the $1\sigma$ scatter. Dashed lines and darker colors correspond to the $\Lambda$CDM simulation dataset, while solid lines and lighter colors denote the observational data.
		\label{fig:2}}
\end{figure}

The median trends of micro-filament mass density contrast distributions are illustrated in Fig.~\ref{fig:2} with respect to their length and thickness. This contrast relative to the background is obtained using Eq.~\ref{eq:1}, substituting the macro-filament density with each micro-filament’s mass density $\rho^{Micro}_{m}$. As shown in Fig.~\ref{fig:2}, all three micro-filament categories ($Total$, $GG$, and $CC$), exhibit similar deviations between the observational and $\Lambda$CDM simulation datasets. Simulated micro-filaments consistently display lower mass density contrast across all shared bins in both length and thickness distributions. In addition, the simulation sample spans broader ranges in length and thickness, except for thickness distribution of the $GG$ category which exhibits a similar range in both the observation and simulation.

Differences between the median trends are clearly pronounced across all three categories. Specifically, $CC$ micro-filaments exhibit lower median mass density contrasts with respect to both length and thickness distributions, while the $GG$ category shows median values that closely align with those of the $Total$ category.

As the GrAviPaSt method identifies micro-filaments based on the gravitational potential of galaxies (Sec.~\ref{sec:2.3}), variations in mass density contrast along both the radial and longitudinal directions are analyzed.

To derive the radial profile, the perpendicular distance ($r$) of each filament galaxy from the corresponding micro-filament skeleton is measured and normalized to the host micro-filament’s radius ($R$). Using these normalized distances $\tau(r/R)$, all filament galaxies are subsequently stacked to construct a comprehensive micro-filament encompassing galaxies at various normalized radial positions. The mass density contrast at different normalized radial distances $\delta_{m}^{\displaystyle \tau}$ is then determined via Eq.~\ref{eq:1}, employing the cumulative mass density at various normalized radii $\rho_{m}^{\displaystyle \tau}$ instead of $\rho^{Macro}_{m}$.

To model and compare the radial profiles across various micro-filament categories within the observational and simulation samples, the generic-$\beta$ model derived from the Plummer profile (\cite{cavaliere1976x,arnaud2009beta,ettori2013mass,galarraga2022relative,yang2025width}) is applied. This profile is defined as:
\begin{equation}
	\label{eq:2}
	\log_{10}(\delta_{m}^{\displaystyle \tau}) = \log_{10}\left(\frac{1+\delta_{m}^1}{(1+\tau^{\alpha})^{\beta}} - 1\right) + \gamma
\end{equation}
where $\delta_{m}^{1}$ denotes the contrast in mass density at $\tau = 1$, and $\alpha$, $\beta$, and $\gamma$ serve as fitting parameters. This profile is selected for its effectiveness in modeling the mass density contrast variation of micro-filaments as a function of normalized radial distance ($\tau$). 

To derive the longitudinal profile, the closest grid point within the micro-filament skeleton to each filament galaxy is first identified. The distance from the starting grid point to this nearest grid point ($l$) is then measured and normalized to the total length of the host micro-filament ($L$), yielding the normalized longitudinal position $\Gamma(l/L)$ for each filament galaxy. Due to the micro-filament identification process employed by the GrAviPaSt method, micro-filaments typically begin near the smaller galaxy system and terminate near the larger system; accordingly, smaller values of $\Gamma$ correspond to regions closer to the smaller system, while larger values reflect proximity to the larger galaxy system. All filament galaxies are subsequently stacked to form a comprehensive micro-filament encompassing galaxies distributed across the full range of normalized longitudinal positions ($\Gamma$). The mass density contrast $\delta_{m}^{\Gamma}$ in each bin spanning 0.1 of the normalized position $\Gamma$, is then computed via Eq.~\ref{eq:1}, using the mass density at various normalized length bins $\rho_{m}^{\Gamma}$ in place of $\rho^{Macro}_{m}$.

Modeling the longitudinal profiles of micro-filaments has been challenging due to their diverse structural variations. In this study, for the first time, an asymmetric Hyperbolic-Secant profile (\cite{johnson1995continuous,fischer2010beta,fischer2025hyperbolic}) as a function of the normalized length ($\Gamma$), is proposed and tested to model and compare the longitudinal profiles across different micro-filament categories ($Total$, $CC$ and $GG$) in both observational and simulation samples. The proposed model is defined as:
\begin{equation}
	\label{eq:3}
	log_{10}(\delta_{m}^{\Gamma}) = \frac{\displaystyle\alpha}{cosh^{\beta+\delta(\Gamma-\Gamma_{\delta_{min}})}\left(\dfrac{\Gamma-\Gamma_{\delta_{min}}}{\gamma}\right)}
\end{equation}
where $\Gamma_{\delta_{\min}}$ denotes the normalized longitudinal position at which $\log_{10}(\delta_{m}^{\Gamma})$ reaches its minimum, and $\alpha$, $\beta$, $\delta$, and $\gamma$ are fitting parameters. Specifically, $\alpha$ acts as the scaling factor; $\beta$ controls the sharpness of the dip; $\delta$ modulates the asymmetry of the profile; $\gamma$ adjusts the width of the profile; and $\Gamma_{\delta_{\min}}$ anchors the location of the minimum.

\begin{figure*}
	\centering
	\includegraphics[width=0.7\textwidth]{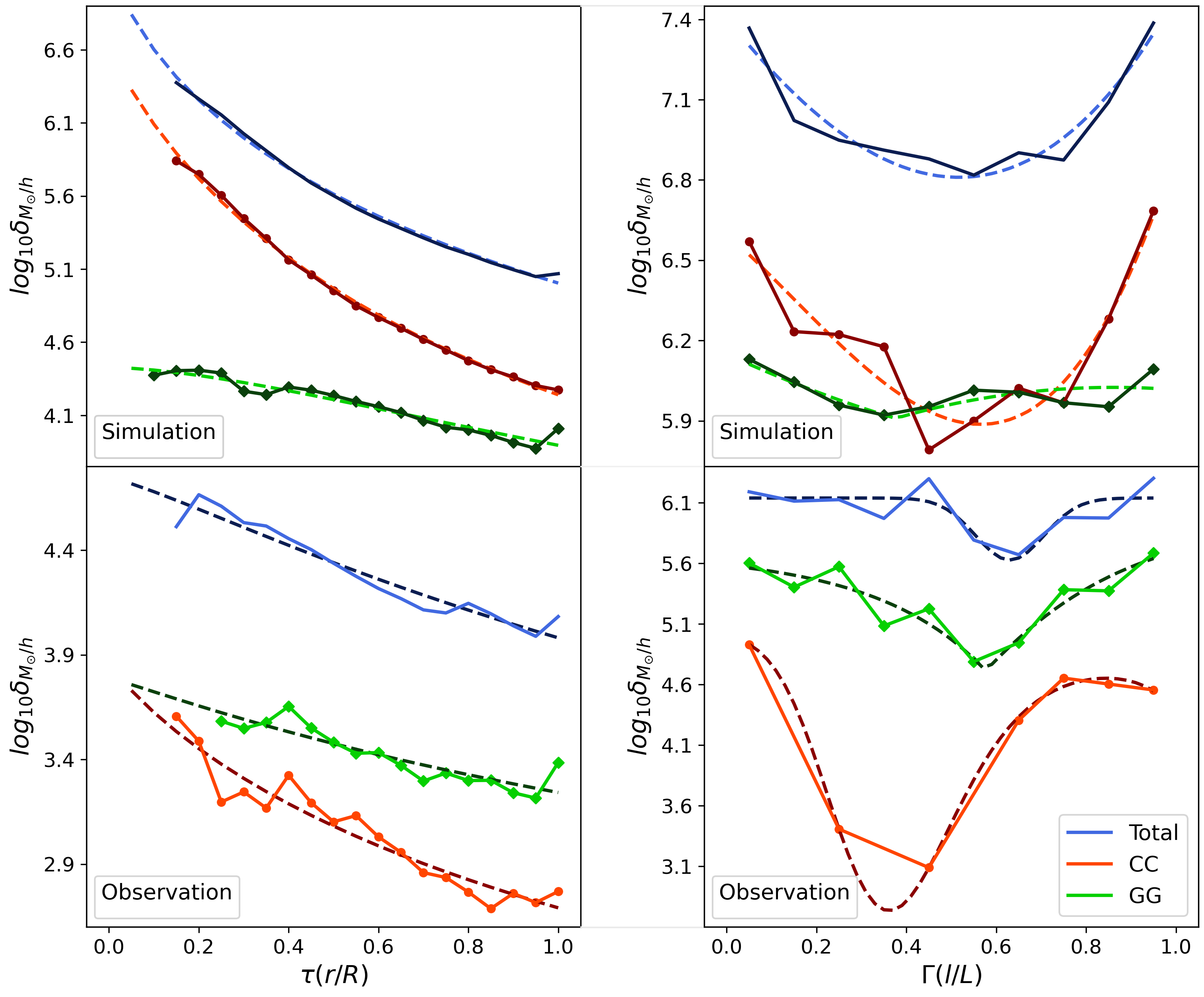}
	\caption{
		Radial (\textit{left}) and longitudinal (\textit{right}) mass density contrast profiles for micro-filament samples are shown for both the $\Lambda$CDM simulation (\textit{top}) and the observational dataset (\textit{bottom}). Each panel presents three micro-filament categories, $Total$, $CC$, and $GG$, distinguished by different colors. Dashed lines represent the fitted curves, based on the Plummer (radial) and Hyperbolic-Secant (longitudinal) profiles.
		\label{fig:3}}
\end{figure*}

The radial and longitudinal profiles of the three micro-filament categories ($Total$, $CC$, and $GG$), for both the observational and $\Lambda$CDM simulation samples are shown in the left and right panels of Fig.~\ref{fig:3}, respectively. The corresponding fit curves are plotted as dashed lines.

\begin{deluxetable*}{cccccccccccccc}  
\tablewidth{0pt}  
\tablecaption{The fitted radial Plummer profiles, as defined in Eq.~\ref{eq:2}, are presented along with the Normalized Root Mean Squared Error ($err$) values corresponding to each fit curve. \label{tab:1}}
\tablehead{  
	\multicolumn{1}{c}{Sample} & \multicolumn{6}{c}{$\Lambda$CDM simulation} & \multicolumn{1}{c}{$ $} & \multicolumn{6}{c}{Observation}\\  
	\colhead{Parameter} & \colhead{$\Delta$} & \colhead{$\alpha$} & \colhead{$\beta$} & \colhead{$\gamma$} & \colhead{$\Delta\delta_{m}^{1}$} & \colhead{$err$} & \colhead{$ $} & \colhead{$\Delta$} & \colhead{$\alpha$} & \colhead{$\beta$} & \colhead{$\gamma$} & \colhead{$\Delta\delta_{m}^{1}$} & \colhead{$err$}
}  
\startdata
Total & 1.41 & 0.73 & 0.67 & 2.19 & 0.06 & 0.017 &  & 0.65 & 1.23 & 0.24 & 0.67 & 0.1 & 0.081 \\
CC & 1.65 & 0.87 & 0.88 & 2.37 & 0.03 & 0.014 &  & 0.84 & 0.88 & 0.54 & 1.09 & 0.08 & 0.083 \\
GG & 0.51 & 1.64 & 0.16 & 0.42 & 0.11 & 0.078 &  & 0.38 & 1.08 & 0.19 & 0.41 & 0.14 & 0.132 \\  
\enddata
\end{deluxetable*}

\begin{deluxetable*}{cccccccccccccc}  
\tablewidth{0pt}  
\tablecaption{The fitted longitudinal Hyperbolic-Secant profiles, as outlined in Eq.~\ref{eq:3}, are summarized along with the Normalized Root Mean Squared Error ($err$) values that serve as indicators of fitting performance. \label{tab:2}}
\tablehead{
	\multicolumn{1}{c}{Sample} & \multicolumn{6}{c}{$\Lambda$CDM simulation} & \multicolumn{1}{c}{$ $} & \multicolumn{6}{c}{Observation}\\  
	\colhead{Parameter} & \colhead{$\Delta$} & \colhead{$\beta$} & \colhead{$\delta$} & \colhead{$\gamma$} & \colhead{$\Delta\Gamma_{\delta_{min}}$} & \colhead{$err$} & \colhead{$ $} & \colhead{$\Delta$} & \colhead{$\beta$} & \colhead{$\delta$} & \colhead{$\gamma$} & \colhead{$\Delta\Gamma_{\delta_{min}}$} & \colhead{$err$}
}  
\startdata
Total & 0.53 & 7.24 & 1.89 & 33.87 & 0.03 & 0.098 &  & 0.49 & 0.73 & -0.97 & 0.03 & 0.04 & 0.173 \\
CC & 0.78 & 7.27 & 5.79 & 20.08 & 0.11 & 0.096 &  & 1.84 & 12.96 & -17.16 & 0.51 & 0.09 & 0.014 \\
GG & 0.19 & 0.0002 & -0.0002 & 0.0008 & 0.02 & 0.187 &  & 0.83 & 0.002 & 0.001 & 0.0007 & 0.03 & 0.124 \\
\enddata
\end{deluxetable*}

As illustrated in Fig.~\ref{fig:3}, the radial mass density contrast consistently decreases with increasing distance from the micro-filament skeleton. The longitudinal profiles exhibit elevated contrast near galaxy systems, descending to a minimum around the filament midpoint. The transition across longitudinal regions varies in shape and intensity, underscoring the diversity in micro-filament profile structures.

Across both radial and longitudinal dimensions, the $Total$, $CC$, and $GG$ categories display largely similar profile behavior. Micro-filaments in the $\Lambda$CDM simulation consistently show higher contrast with respect to the background than their observational counterparts. Notably, in the simulation sample, the $CC$ category often demonstrates a higher mass density contrast than $GG$, whereas in the observational sample, $GG$ micro-filaments display stronger contrast relative to $CC$. This divergence in profile shapes between $CC$ and $GG$ is more pronounced in the simulation than in the observational data. Furthermore, the $CC$ category in the simulation more closely resembles the $Total$ profile, while in the observational data, it is the $GG$ category that exhibits more similarity to the $Total$.

Tab.~\ref{tab:1} and Tab.~\ref{tab:2} summarize the results of the fitted radial and longitudinal profile curves, along with their associated Normalized Root Mean Squared Error (NRMSE) values. The NRMSE is computed by first evaluating the root mean squared error (RMSE) between the fitted profile and the corresponding data points, and then normalizing this value by the range of the data, defined as the difference between its maximum and minimum values. This normalization facilitates comparison across different filament categories and profile types by removing scale dependence.
In addition to these results, the fitted values of the $\alpha$ parameter for the longitudinal profiles are as follows: for the simulation sample, $\alpha = -815.13$, $-448.91$, and $-1.83$ for the $Total$, $CC$, and $GG$ micro-filaments respectively; and for the observational sample, $\alpha = -0.72$, $-2.29$, and $-1.19$ for the same categories. Except for $\alpha$, which acts as the scale factor, the mass density contrast span (hereafter denoted as span $\Delta$) of each fitted curve is presented in Tab.~\ref{tab:2} for comparative purposes. As shown in these tables and the Fig.~\ref{fig:3}, the span of the radial and longitudinal profiles differs between the $\Lambda$CDM simulation and the observational sample. The radial profiles of the simulation micro-filaments exhibit wider spans across all three categories compared to their observational counterparts. For the longitudinal profiles, the $Total$ micro-filaments in both samples show similar spans, while the $CC$ and $GG$ micro-filaments in the observational sample display wider spans than those in the simulation.

Additionally, the differences between the mass density contrast as measured from the data and indicated by the fitted profiles at normalized radius $\tau = 1$ and normalized length $\Gamma = \Gamma_{\delta_{min}}$ are presented as $\Delta\delta_{m}^{1}$ and $\Delta\Gamma_{\delta_{min}}$ in Tab.~\ref{tab:1} and Tab.~\ref{tab:2} respectively. The value of $\Delta\Gamma_{\delta_{min}}$ is correlated with $\delta$, and both serve as indicators of the asymmetry in the longitudinal profiles of micro-filaments. The magnitude of $\delta$ reflects the degree of asymmetry, while its sign conveys the direction of deviation. Positive $\delta$ values correspond to asymmetry toward the bigger galaxy system, and negative values toward the smaller one.

The $GG$ micro-filaments show minimal asymmetry in both simulation and observational samples, likely due to the limited richness variation among the galaxy system pairs in this category (see Sec.~\ref{sec:2.3.2}). In contrast, the $CC$ category exhibits the most pronounced asymmetry, possibly resulting from the wide richness range of its galaxy system pairs. The $Total$ sample in both simulation and observation shows mild asymmetry, which may be attributed to the large number of micro-filaments and the broad diversity of galaxy system pairs.

These structural differences between the simulation and observational micro-filament catalogs reveal disparities between the $\Lambda$CDM model’s representation of filamentary structures and their observed counterparts.

\subsection{Filament Galaxies\label{sec:3.3}}

To further explore the differences between $CC$ and $GG$ micro-filaments, various properties of their associated galaxies, such as distance to the host micro-filament skeleton, color, stellar mass, and star formation activity are analyzed.

\begin{figure}
	\centering
	\includegraphics[width=0.43\textwidth]{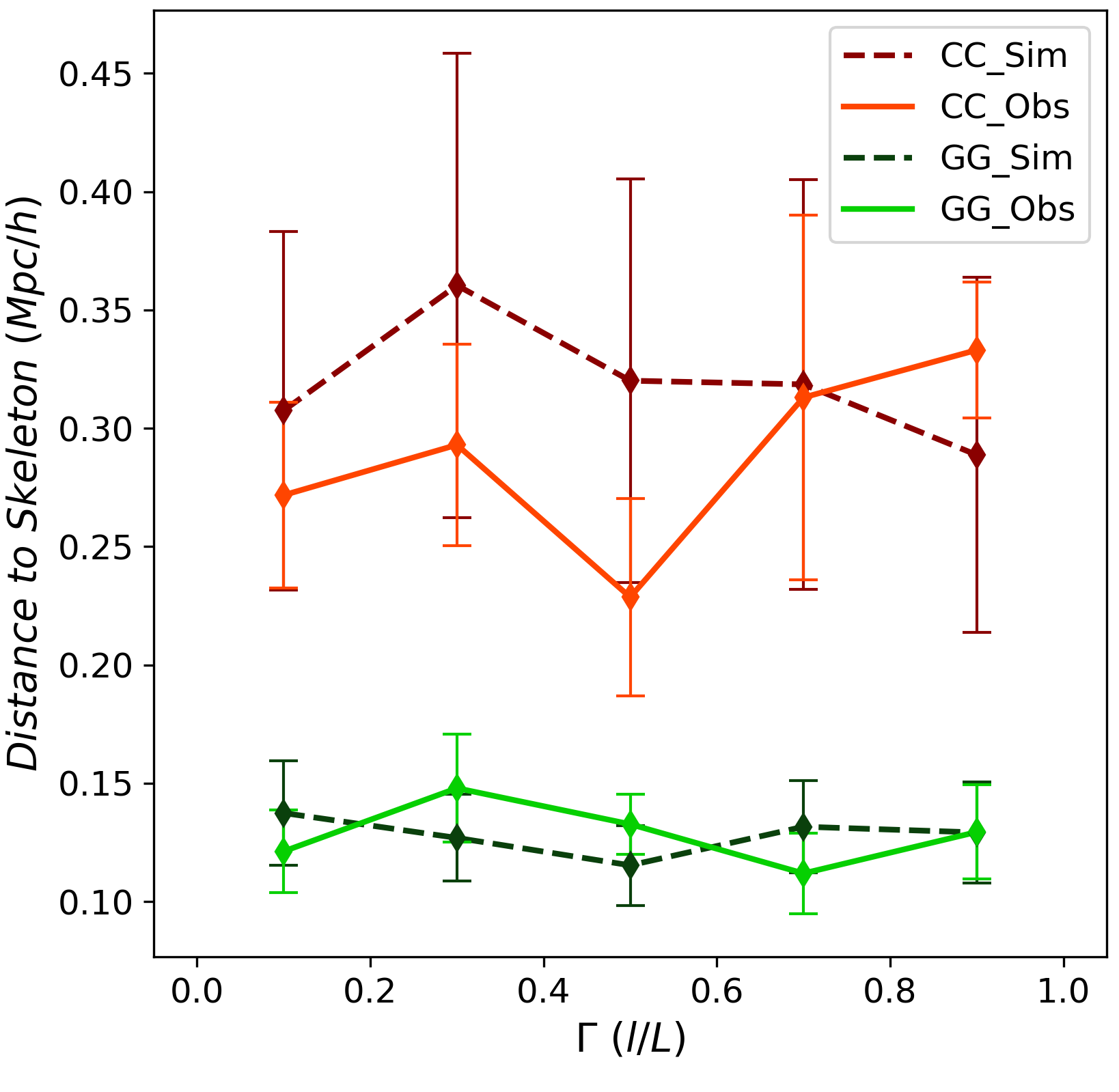}
	\caption{
		Median trends for distances to micro-filament skeletons are shown relative to the normalized longitudinal position ($\Gamma$), along with corresponding 1$\sigma$ errors, for $CC$ and $GG$ micro-filaments. Results are presented for both the $\Lambda$CDM simulation and observational datasets, with dashed lines indicating simulation-based medians and solid lines representing observational values.
		\label{fig:4}}
\end{figure}

The distance of filament galaxies to the skeleton, measured along the normalized micro-filament length ($\Gamma$), is illustrated in Fig.~\ref{fig:4}. Across both simulation and observational datasets, $CC$ galaxies consistently reside farther from the skeleton compared to $GG$ galaxies. This pattern is consistent with the broader transverse extent of $CC$ micro-filaments as illustrated in the right column of Fig.~\ref{fig:2}. While the difference between simulation and observation is more pronounced in the $CC$ category than in $GG$, the $\Lambda$CDM reconstruction of micro-filaments exhibits only subtle deviations from observations with respect to the distance of filament galaxies to skeleton along the length of micro-filaments.

\begin{figure}
	\centering
	\includegraphics[width=0.47\textwidth]{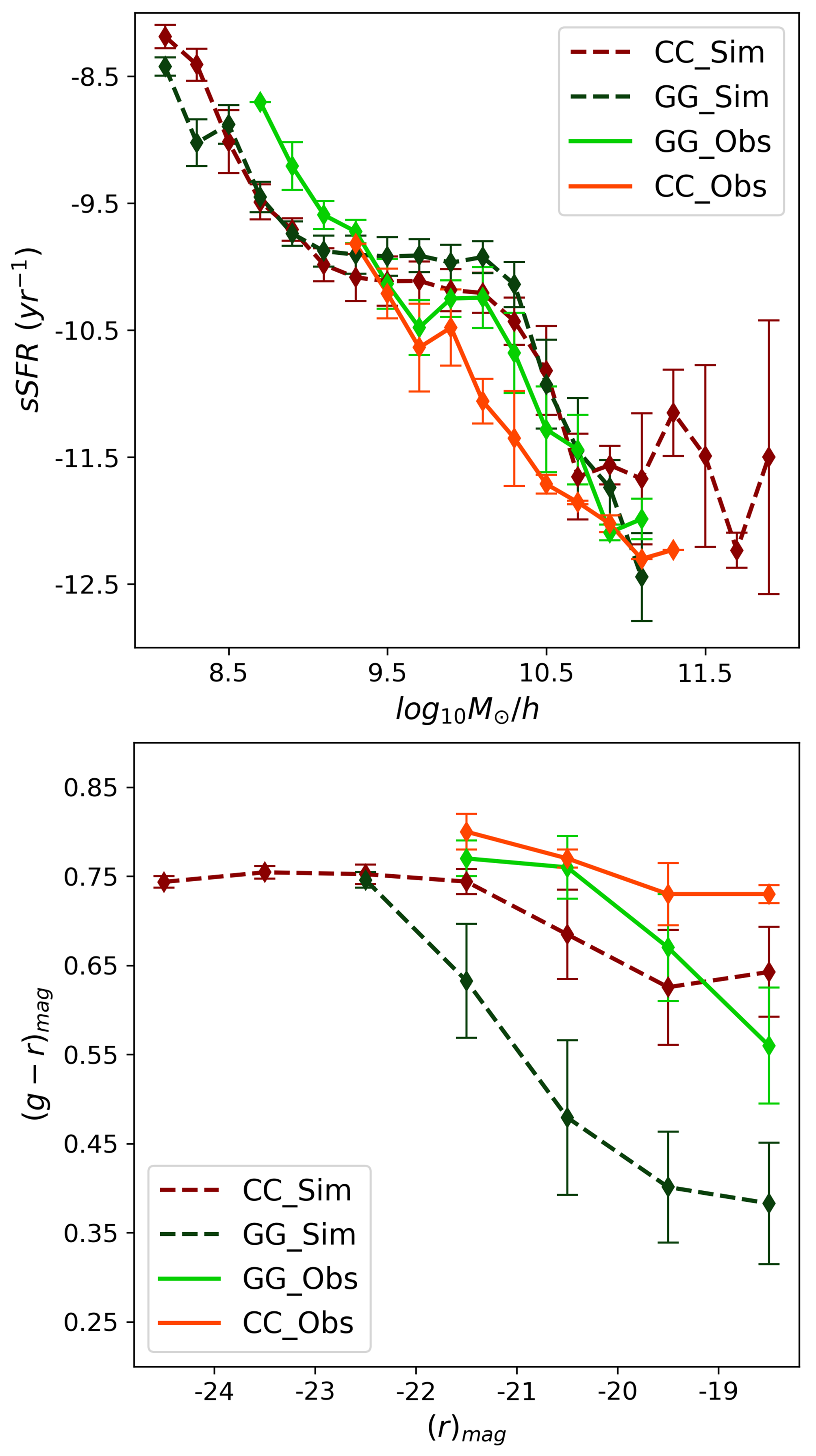}
	\caption{
		Median trends and associated $1\sigma$ errors of the specific star formation rate ($sSFR$) of filament galaxies with respect to stellar mass ($top$) and color-magnitude diagram ($bottom$), shown for both $\Lambda$CDM simulation (dashed lines) and observational (solid lines) samples. The $CC$ and $GG$ filament categories are represented using dark and light colors, respectively.
		\label{fig:5}}
\end{figure}

In terms of stellar mass, and as illustrated in the top panel of Fig.~\ref{fig:5}, filament galaxies in the TNG300-1 simulation exhibit a broader mass range compared to those in the SDSS sample across both the $CC$ and $GG$ categories. The $GG$ galaxies in TNG300-1 cover low to intermediate masses, while its $CC$ galaxies extend from low to high masses. Conversely, SDSS galaxies in both categories mostly have intermediate stellar masses.

Across the stellar mass range of $\log_{10}(M_{\odot}/h) \sim 9$–$11$, both the simulation and observation samples contain $CC$ and $GG$ galaxies. Notably, the $GG$ filament galaxies in TNG300-1 exhibit the highest level of star-formation in this mass range, whereas the $CC$ galaxies in SDSS show the least. In aggregate, filament galaxies reconstructed in TNG300-1 show higher star-formation activity than their observational counterparts.

Regarding photometric properties, the lower panel of Fig.~\ref{fig:5} indicates that TNG300-1 filament galaxies cover a broader range of r-band magnitudes than those observed in SDSS. For galaxies fainter than approximately $-22$, both $CC$ and $GG$ categories are present in simulation and observational samples. Notabely, SDSS filament galaxies generally host older stellar populations across both categories, in contrast to the younger populations found in TNG300-1.

These results may further emphasize the discrepancies between the baryonic properties of galaxies in the TNG300-1 simulation and those observed in SDSS data. While comparisons between IllustrisTNG and observational data have previously been explored for missing baryons in different environments (e.g., \cite{martizzi2019baryons}), feedback mechanisms (e.g., \cite{donnari2021quenched, piotrowska2022quenching}), galaxy clusters (e.g., \cite{miller2025differences}), etc., this study presents the first investigation of discrepancies in filament galaxies within the IllustrisTNG300-1 simulation with observations.

\section{Summary and Conclusion\label{sec:4}}

In this study, an extended comparative analysis of micro-filament structural characteristics is conducted between the SDSS DR10 observational dataset ($z < 0.05$; $M_r \leq -18$) and the TNG300-1 $\Lambda$CDM simulation sample, both selected under consistent criteria.

Using the novel GrAviPaSt filament-finder method, comprehensive catalogs of macro- and micro-filaments are constructed for both datasets. Throughout this investigation, in addition to the $Total$ sample, two distinct micro-filament subclasses are examined: $CC$ (cluster-to-cluster) and $GG$ (group-to-group) filaments.

The comparison between simulation and observational cosmic filaments yields several key findings:

(I) Although the $\Lambda$CDM model produces extended macro-filaments, such structures are largely absent in SDSS. When present, observed macro-filaments exhibit higher average mass densities than those predicted by the simulation.

(II) At fixed lengths and radii, micro-filaments in the $\Lambda$CDM model display significantly lower mass densities compared to their observational counterparts, indicating a structural discrepancy.

(III) Radial profiles from the $\Lambda$CDM micro-filament catalog suggest denser filamentary structures in the simulation than in the SDSS sample. Notably, $CC$ type micro-filaments appear denser than $GG$ type within the simulation, whereas SDSS observations exhibit the opposite trend. These findings were further confirmed by the longitudinal profiles of micro-filaments.

(IV) Galaxies within SDSS micro-filaments, particularly of intermediate stellar mass, tend to exhibit lower specific star formation rates (sSFRs) and older stellar populations. In contrast, the TNG300-1 filaments host more actively star-forming galaxies in the same mass range.

These results highlight persistent tensions between the $\Lambda$CDM-based IllustrisTNG300-1 simulation and observational data from SDSS, in both the morphology of cosmic filaments and the evolutionary properties of filament galaxies. This underscores potential gaps in our understanding of the environmental processes that govern cosmic filaments. It is worth noting that the presented discrepancy requires further investigation to assess the dependence of these outcomes on methodological choices, such as structure identification techniques and the selection of simulation and observational datasets. Such investigation could employ varied filament finder methods, incorporating different structure identification procedures and filament definitions (e.g., \cite{bonnaire2020t,pfeifer2022cows}). Moreover, variations in the criteria used to classify groups and clusters may influence the resulting filament catalogs. These methodological choices may affect the detection efficiency, characteristic length scales, and density of filaments, thereby contributing to the observed mismatches. In addition, comparison between simulation and observational data across different $\Lambda$CDM-based cosmological simulations can be conducted, that employ various semi-analytical and hydrodynamical models for reconstructing baryonic properties (e.g., \cite{schaye2015eagle,ayromlou2021galaxy}), and across multiple galaxy redshift surveys (e.g., \cite{colless20012df,adame2024early}).

\bibliography{Filament_Rift}{}

\begin{thebibliography}{}
\expandafter\ifx\csname natexlab\endcsname\relax\def\natexlab#1{#1}\fi
\providecommand{\url}[1]{\href{#1}{#1}}
\providecommand{\dodoi}[1]{doi:~\href{http://doi.org/#1}{\nolinkurl{#1}}}
\providecommand{\doeprint}[1]{\href{http://ascl.net/#1}{\nolinkurl{http://ascl.net/#1}}}
\providecommand{\doarXiv}[1]{\href{https://arxiv.org/abs/#1}{\nolinkurl{https://arxiv.org/abs/#1}}}

\bibitem[{A. Adame {et~al.}(2024)Adame, Aguilar, Ahlen, Alam, Aldering,
  Alexander, Alfarsy, Prieto, Alvarez, Alves, {et~al.}}]{adame2024early}
Adame, A., Aguilar, J., Ahlen, S., {et~al.} 2024, \bibinfo{title}{The early
  data release of the dark energy spectroscopic instrument,} The Astronomical
  Journal, 168, 58

\bibitem[{P.~A. Ade {et~al.}(2016)Ade, Aghanim, Arnaud, Ashdown, Aumont,
  Baccigalupi, Banday, Barreiro, Bartlett, Bartolo, {et~al.}}]{ade2016planck}
Ade, P.~A., Aghanim, N., Arnaud, M., {et~al.} 2016, \bibinfo{title}{Planck 2015
  results-xiii. cosmological parameters,} Astronomy \& Astrophysics, 594, A13

\bibitem[{N. Aghanim {et~al.}(2020)Aghanim, Akrami, Ashdown, Aumont,
  Baccigalupi, Ballardini, Banday, Barreiro, Bartolo, Basak,
  {et~al.}}]{aghanim2020planck}
Aghanim, N., Akrami, Y., Ashdown, M., {et~al.} 2020, \bibinfo{title}{Planck
  2018 results-V. CMB power spectra and likelihoods,} Astronomy \&
  Astrophysics, 641, A5

\bibitem[{C.~P. Ahn {et~al.}(2014)Ahn, Alexandroff, Prieto, Anders, Anderson,
  Anderton, Andrews, Aubourg, Bailey, Bastien, {et~al.}}]{ahn2014tenth}
Ahn, C.~P., Alexandroff, R., Prieto, C.~A., {et~al.} 2014, \bibinfo{title}{The
  tenth data release of the sloan digital sky survey: First spectroscopic data
  from the sdss-iii apache point observatory galactic evolution experiment,}
  The Astrophysical Journal Supplement Series, 211, 17

\bibitem[{M.~A. Aragon-Calvo {et~al.}(2019)Aragon-Calvo, Neyrinck, \&
  Silk}]{aragon2019galaxy}
Aragon-Calvo, M.~A., Neyrinck, M.~C., \& Silk, J. 2019, \bibinfo{title}{Galaxy
  Quenching from Cosmic Web Detachment,} The Astrophysical Journal, 2, 7

\bibitem[{M. Arnaud(2009)Arnaud}]{arnaud2009beta}
Arnaud, M. 2009, \bibinfo{title}{The $\beta$-model of the intracluster
  medium-commentary on: Cavaliere a. and fusco-femiano r., 1976, a\&a, 49,
  137,} Astronomy \& Astrophysics, 500, 103

\bibitem[{M. Ayromlou {et~al.}(2021)Ayromlou, Kauffmann, Yates, Nelson, \&
  White}]{ayromlou2021galaxy}
Ayromlou, M., Kauffmann, G., Yates, R.~M., Nelson, D., \& White, S.~D. 2021,
  \bibinfo{title}{Galaxy formation with L-GALAXIES: modelling the environmental
  dependency of galaxy evolution and comparing with observations,} Monthly
  Notices of the Royal Astronomical Society, 505, 492

\bibitem[{M.~R. Blanton {et~al.}(2005)Blanton, Schlegel, Strauss, Brinkmann,
  Finkbeiner, Fukugita, Gunn, Hogg, Ivezi{\'c}, Knapp,
  {et~al.}}]{blanton2005new}
Blanton, M.~R., Schlegel, D.~J., Strauss, M.~A., {et~al.} 2005,
  \bibinfo{title}{New York University Value-Added Galaxy Catalog: a galaxy
  catalog based on new public surveys,} The Astronomical Journal, 129, 2562

\bibitem[{J.~R. Bond {et~al.}(1996)Bond, Kofman, \&
  Pogosyan}]{bond1996filaments}
Bond, J.~R., Kofman, L., \& Pogosyan, D. 1996, \bibinfo{title}{How filaments of
  galaxies are woven into the cosmic web,} Nature, 380, 603

\bibitem[{T. Bonnaire {et~al.}(2020)Bonnaire, Aghanim, Decelle, \&
  Douspis}]{bonnaire2020t}
Bonnaire, T., Aghanim, N., Decelle, A., \& Douspis, M. 2020,
  \bibinfo{title}{T-ReX: a graph-based filament detection method,} Astronomy \&
  Astrophysics, 637, A18

\bibitem[{J. Brinchmann {et~al.}(2004)Brinchmann, Charlot, White, Tremonti,
  Kauffmann, Heckman, \& Brinkmann}]{brinchmann2004physical}
Brinchmann, J., Charlot, S., White, S.~D., {et~al.} 2004, \bibinfo{title}{The
  physical properties of star-forming galaxies in the low-redshift Universe,}
  Monthly notices of the royal astronomical society, 351, 1151

\bibitem[{M. Cautun {et~al.}(2014)Cautun, Van De~Weygaert, Jones, \&
  Frenk}]{cautun2014evolution}
Cautun, M., Van De~Weygaert, R., Jones, B.~J., \& Frenk, C.~S. 2014,
  \bibinfo{title}{Evolution of the cosmic web,} Monthly Notices of the Royal
  Astronomical Society, 441, 2923

\bibitem[{A. Cavaliere \& R. Fusco-Femiano(1976)Cavaliere \&
  Fusco-Femiano}]{cavaliere1976x}
Cavaliere, A., \& Fusco-Femiano, R. 1976, \bibinfo{title}{X-rays from hot
  plasma in clusters of galaxies,} Astronomy and Astrophysics, vol. 49, no. 1,
  May 1976, p. 137-144., 49, 137

\bibitem[{M. Colless {et~al.}(2001)Colless, Dalton, Maddox, Sutherland,
  Norberg, Cole, Bland-Hawthorn, Bridges, Cannon, Collins,
  {et~al.}}]{colless20012df}
Colless, M., Dalton, G., Maddox, S., {et~al.} 2001, \bibinfo{title}{The 2df
  galaxy redshift survey: spectra and redshifts,} Monthly Notices of the Royal
  Astronomical Society, 328, 1039

\bibitem[{M. Davis {et~al.}(1982)Davis, Huchra, Latham, \&
  Tonry}]{davis1982survey}
Davis, M., Huchra, J., Latham, D.~W., \& Tonry, J. 1982, \bibinfo{title}{A
  survey of galaxy redshifts. II-The large scale space distribution,}
  Astrophysical Journal, Part 1, vol. 253, Feb. 15, 1982, p. 423-445., 253, 423

\bibitem[{M. Donnari {et~al.}(2021)Donnari, Pillepich, Nelson, Marinacci,
  Vogelsberger, \& Hernquist}]{donnari2021quenched}
Donnari, M., Pillepich, A., Nelson, D., {et~al.} 2021, \bibinfo{title}{Quenched
  fractions in the IllustrisTNG simulations: comparison with observations and
  other theoretical models,} Monthly Notices of the Royal Astronomical Society,
  506, 4760

\bibitem[{S. Ettori {et~al.}(2013)Ettori, Donnarumma, Pointecouteau, Reiprich,
  Giodini, Lovisari, \& Schmidt}]{ettori2013mass}
Ettori, S., Donnarumma, A., Pointecouteau, E., {et~al.} 2013,
  \bibinfo{title}{Mass profiles of galaxy clusters from X-ray analysis,} Space
  Science Reviews, 177, 119

\bibitem[{M. Fischer(2025)Fischer}]{fischer2025hyperbolic}
Fischer, M. 2025, in International encyclopedia of statistical science
  (Springer), 1151--1152

\bibitem[{M.~J. Fischer \& D. Vaughan(2010)Fischer \&
  Vaughan}]{fischer2010beta}
Fischer, M.~J., \& Vaughan, D. 2010, \bibinfo{title}{The beta-hyperbolic secant
  distribution,} Austrian Journal of Statistics, 39, 245

\bibitem[{D. Gal{\'a}rraga-Espinosa {et~al.}(2022)Gal{\'a}rraga-Espinosa,
  Langer, \& Aghanim}]{galarraga2022relative}
Gal{\'a}rraga-Espinosa, D., Langer, M., \& Aghanim, N. 2022,
  \bibinfo{title}{Relative distribution of dark matter, gas, and stars around
  cosmic filaments in the IllustrisTNG simulation,} Astronomy \& Astrophysics,
  661, A115

\bibitem[{D. Gal{\'a}rraga-Espinosa {et~al.}(2024)Gal{\'a}rraga-Espinosa,
  Cadiou, Gouin, White, Springel, Pakmor, Hadzhiyska, Bose, Ferlito, Hernquist,
  {et~al.}}]{galarraga2024evolution}
Gal{\'a}rraga-Espinosa, D., Cadiou, C., Gouin, C., {et~al.} 2024,
  \bibinfo{title}{Evolution of cosmic filaments in the MTNG simulation,}
  Astronomy \& Astrophysics, 684, A63

\bibitem[{P. Ghafour \& S. Tavasoli(2025)Ghafour \&
  Tavasoli}]{ghafour2025gravipast}
Ghafour, P., \& Tavasoli, S. 2025, \bibinfo{title}{GrAviPaSt's Lens to the
  Past: Unveiling the Evolution of Filamentary Structures,} The Astronomical
  Journal, 990

\bibitem[{F. Habibi {et~al.}(2018)Habibi, Baghram, \&
  Tavasoli}]{habibi2018peculiar}
Habibi, F., Baghram, S., \& Tavasoli, S. 2018, \bibinfo{title}{Peculiar
  velocity measurement in a clumpy universe,} International Journal of Modern
  Physics D, 27, 1850019

\bibitem[{A. He {et~al.}(2023)He, Ivanov, An, \& Gluscevic}]{he2023s8}
He, A., Ivanov, M.~M., An, R., \& Gluscevic, V. 2023, \bibinfo{title}{S8
  tension in the context of dark matter--baryon scattering,} The Astrophysical
  Journal Letters, 954, L8

\bibitem[{H. Hildebrandt {et~al.}(2017)Hildebrandt, Viola, Heymans, Joudaki,
  Kuijken, Blake, Erben, Joachimi, Klaes, Miller,
  {et~al.}}]{hildebrandt2017kids}
Hildebrandt, H., Viola, M., Heymans, C., {et~al.} 2017,
  \bibinfo{title}{KiDS-450: cosmological parameter constraints from tomographic
  weak gravitational lensing,} Monthly Notices of the Royal Astronomical
  Society, 465, 1454

\bibitem[{N.~L. Johnson {et~al.}(1995)Johnson, Kotz, \&
  Balakrishnan}]{johnson1995continuous}
Johnson, N.~L., Kotz, S., \& Balakrishnan, N. 1995, Continuous univariate
  distributions, volume 2, Vol.~2 (John wiley \& sons)

\bibitem[{A. Kashlinsky {et~al.}(2008)Kashlinsky, Atrio-Barandela, Kocevski, \&
  Ebeling}]{kashlinsky2008measurement}
Kashlinsky, A., Atrio-Barandela, F., Kocevski, D., \& Ebeling, H. 2008,
  \bibinfo{title}{A measurement of large-scale peculiar velocities of clusters
  of galaxies: results and cosmological implications,} The Astrophysical
  Journal, 686, L49

\bibitem[{G. Kauffmann {et~al.}(2003)Kauffmann, Heckman, White, Charlot,
  Tremonti, Brinchmann, Bruzual, Peng, Seibert, Bernardi,
  {et~al.}}]{kauffmann2003stellar}
Kauffmann, G., Heckman, T.~M., White, S.~D., {et~al.} 2003,
  \bibinfo{title}{Stellar masses and star formation histories for 105 galaxies
  from the Sloan Digital Sky Survey,} Monthly Notices of the Royal Astronomical
  Society, 341, 33

\bibitem[{A. Klypin {et~al.}(1999)Klypin, Kravtsov, Valenzuela, \&
  Prada}]{klypin1999missing}
Klypin, A., Kravtsov, A.~V., Valenzuela, O., \& Prada, F. 1999,
  \bibinfo{title}{Where are the missing galactic satellites?} The Astrophysical
  Journal, 522, 82

\bibitem[{A.~V. Kravtsov \& S. Borgani(2012)Kravtsov \&
  Borgani}]{kravtsov2012formation}
Kravtsov, A.~V., \& Borgani, S. 2012, \bibinfo{title}{Formation of galaxy
  clusters,} Annual Review of Astronomy and Astrophysics, 50, 353

\bibitem[{N. Malavasi {et~al.}(2020)Malavasi, Aghanim, Douspis, Tanimura, \&
  Bonjean}]{malavasi2020characterising}
Malavasi, N., Aghanim, N., Douspis, M., Tanimura, H., \& Bonjean, V. 2020,
  \bibinfo{title}{Characterising filaments in the SDSS volume from the galaxy
  distribution,} Astronomy \& Astrophysics, 642, A19

\bibitem[{D. Martizzi {et~al.}(2019)Martizzi, Vogelsberger, Artale, Haider,
  Torrey, Marinacci, Nelson, Pillepich, Weinberger, Hernquist,
  {et~al.}}]{martizzi2019baryons}
Martizzi, D., Vogelsberger, M., Artale, M.~C., {et~al.} 2019,
  \bibinfo{title}{Baryons in the Cosmic Web of IllustrisTNG--I: gas in knots,
  filaments, sheets, and voids,} Monthly Notices of the Royal Astronomical
  Society, 486, 3766

\bibitem[{D. Miller {et~al.}(2025)Miller, Pallero, Tissera, \&
  Bla{\~n}a}]{miller2025differences}
Miller, D., Pallero, D., Tissera, P.~B., \& Bla{\~n}a, M. 2025,
  \bibinfo{title}{Differences in baryonic and dark matter scaling relations of
  galaxy clusters: A comparison between IllustrisTNG simulations and
  observations,} Astronomy \& Astrophysics, 698, A237

\bibitem[{D. Nelson {et~al.}(2019)Nelson, Springel, Pillepich, Rodriguez-Gomez,
  Torrey, Genel, Vogelsberger, Pakmor, Marinacci, Weinberger,
  {et~al.}}]{nelson2019illustristng}
Nelson, D., Springel, V., Pillepich, A., {et~al.} 2019, \bibinfo{title}{The
  IllustrisTNG simulations: public data release,} Computational Astrophysics
  and Cosmology, 6, 1

\bibitem[{P. Peebles(1967)Peebles}]{peebles1967gravitational}
Peebles, P. 1967, \bibinfo{title}{The gravitational instability of the
  universe,} Astrophysical Journal, vol. 147, p. 859, 147, 859

\bibitem[{P. Peebles(2001)Peebles}]{peebles2001void}
Peebles, P. 2001, \bibinfo{title}{The void phenomenon,} The Astrophysical
  Journal, 557, 495

\bibitem[{P.~J.~E. Peebles(2020)Peebles}]{peebles2020large}
Peebles, P. J.~E. 2020, The large-scale structure of the universe, Vol.~96
  (Princeton university press)

\bibitem[{S. Pfeifer {et~al.}(2022)Pfeifer, Libeskind, Hoffman, Hellwing,
  Bilicki, \& Naidoo}]{pfeifer2022cows}
Pfeifer, S., Libeskind, N.~I., Hoffman, Y., {et~al.} 2022,
  \bibinfo{title}{COWS: a filament finder for Hessian cosmic web identifiers,}
  Monthly Notices of the Royal Astronomical Society, 514, 470

\bibitem[{J.~M. Piotrowska {et~al.}(2022)Piotrowska, Bluck, Maiolino, \&
  Peng}]{piotrowska2022quenching}
Piotrowska, J.~M., Bluck, A.~F., Maiolino, R., \& Peng, Y. 2022,
  \bibinfo{title}{On the quenching of star formation in observed and simulated
  central galaxies: evidence for the role of integrated AGN feedback,} Monthly
  Notices of the Royal Astronomical Society, 512, 1052

\bibitem[{S. Planelles {et~al.}(2014)Planelles, Borgani, Fabjan, Killedar,
  Murante, Granato, Ragone-Figueroa, \& Dolag}]{planelles2014role}
Planelles, S., Borgani, S., Fabjan, D., {et~al.} 2014, \bibinfo{title}{On the
  role of AGN feedback on the thermal and chemodynamical properties of the hot
  intracluster medium,} Monthly Notices of the Royal Astronomical Society, 438,
  195

\bibitem[{R.~C. Prim(1957)Prim}]{prim1957shortest}
Prim, R.~C. 1957, \bibinfo{title}{Shortest connection networks and some
  generalizations,} The Bell System Technical Journal, 36, 1389

\bibitem[{M.~A. Sabogal {et~al.}(2024)Sabogal, Silva, Nunes, Kumar,
  Di~Valentino, \& Giar{\`e}}]{sabogal2024quantifying}
Sabogal, M.~A., Silva, E., Nunes, R.~C., {et~al.} 2024,
  \bibinfo{title}{Quantifying the S 8 tension and evidence for interacting dark
  energy from redshift-space distortion measurements,} Physical Review D, 110,
  123508

\bibitem[{S. Salim {et~al.}(2007)Salim, Rich, Charlot, Brinchmann, Johnson,
  Schiminovich, Seibert, Mallery, Heckman, Forster, {et~al.}}]{salim2007uv}
Salim, S., Rich, R.~M., Charlot, S., {et~al.} 2007, \bibinfo{title}{UV star
  formation rates in the local universe,} The Astrophysical Journal Supplement
  Series, 173, 267

\bibitem[{J. Schaye {et~al.}(2015)Schaye, Crain, Bower, Furlong, Schaller,
  Theuns, Dalla~Vecchia, Frenk, McCarthy, Helly, {et~al.}}]{schaye2015eagle}
Schaye, J., Crain, R.~A., Bower, R.~G., {et~al.} 2015, \bibinfo{title}{The
  EAGLE project: simulating the evolution and assembly of galaxies and their
  environments,} Monthly Notices of the Royal Astronomical Society, 446, 521

\bibitem[{A. Singh {et~al.}(2020)Singh, Mahajan, \& Bagla}]{singh2020study}
Singh, A., Mahajan, S., \& Bagla, J.~S. 2020, \bibinfo{title}{Study of galaxies
  on large-scale filaments in simulations,} Monthly Notices of the Royal
  Astronomical Society, 497, 2265

\bibitem[{V. Springel(2010)Springel}]{springel2010pur}
Springel, V. 2010, \bibinfo{title}{E pur si muove: Galilean-invariant
  cosmological hydrodynamical simulations on a moving mesh,} Monthly Notices of
  the Royal Astronomical Society, 401, 791

\bibitem[{V. Springel {et~al.}(2005)Springel, White, Jenkins, Frenk, Yoshida,
  Gao, Navarro, Thacker, Croton, Helly, {et~al.}}]{springel2005simulations}
Springel, V., White, S.~D., Jenkins, A., {et~al.} 2005,
  \bibinfo{title}{Simulations of the formation, evolution and clustering of
  galaxies and quasars,} nature, 435, 629

\bibitem[{C. Stoughton {et~al.}(2002)Stoughton, Lupton, Bernardi, Blanton,
  Burles, Castander, Connolly, Eisenstein, Frieman, Hennessy,
  {et~al.}}]{stoughton2002sloan}
Stoughton, C., Lupton, R.~H., Bernardi, M., {et~al.} 2002,
  \bibinfo{title}{Sloan digital sky survey: early data release,} The
  Astronomical Journal, 123, 485

\bibitem[{S. Tavasoli(2021)Tavasoli}]{tavasoli2021void}
Tavasoli, S. 2021, \bibinfo{title}{Void Galaxy Distribution: A Challenge for
  $\Lambda$CDM,} The Astrophysical Journal Letters, 916, L24

\bibitem[{S. Tavasoli {et~al.}(2015)Tavasoli, Rahmani, Khosroshahi, Vasei, \&
  Lehnert}]{tavasoli2015void}
Tavasoli, S., Rahmani, H., Khosroshahi, H.~G., Vasei, K., \& Lehnert, M.~D.
  2015, \bibinfo{title}{The Galaxy Population In Voids: Are All Voids The
  Same?} The Astrophysical Journal Letters, 803, L23

\bibitem[{S. Tavasoli {et~al.}(2013)Tavasoli, Vasei, \&
  Mohayaee}]{tavasoli2013challenge}
Tavasoli, S., Vasei, K., \& Mohayaee, R. 2013, \bibinfo{title}{The challenge of
  large and empty voids in the SDSS DR7 redshift survey,} Astronomy \&
  Astrophysics, 553, A15

\bibitem[{E. Tempel {et~al.}(2014)Tempel, Tamm, Gramann, Tuvikene,
  Liivam{\"a}gi, Suhhonenko, Kipper, Einasto, \& Saar}]{tempel2014flux}
Tempel, E., Tamm, A., Gramann, M., {et~al.} 2014, \bibinfo{title}{Flux-and
  volume-limited groups/clusters for the SDSS galaxies: catalogues and mass
  estimation,} Astronomy \& Astrophysics, 566, A1

\bibitem[{T. Vernstrom {et~al.}(2021)Vernstrom, Heald, Vazza, Galvin, West,
  Locatelli, Fornengo, \& Pinetti}]{vernstrom2021discovery}
Vernstrom, T., Heald, G., Vazza, F., {et~al.} 2021, \bibinfo{title}{Discovery
  of magnetic fields along stacked cosmic filaments as revealed by radio and
  X-ray emission,} Monthly Notices of the Royal Astronomical Society, 505, 4178

\bibitem[{M. Vogelsberger {et~al.}(2014)Vogelsberger, Genel, Springel, Torrey,
  Sijacki, Xu, Snyder, Bird, Nelson, \& Hernquist}]{vogelsberger2014properties}
Vogelsberger, M., Genel, S., Springel, V., {et~al.} 2014,
  \bibinfo{title}{Properties of galaxies reproduced by a hydrodynamic
  simulation,} Nature, 509, 177

\bibitem[{B. Vulcani {et~al.}(2019)Vulcani, Poggianti, Moretti, Gullieuszik,
  Fritz, Franchetto, Fasano, Bettoni, \& Jaff{\'e}}]{vulcani2019gasp}
Vulcani, B., Poggianti, B.~M., Moretti, A., {et~al.} 2019,
  \bibinfo{title}{GASP--XVI. Does cosmic web enhancement turn on star formation
  in galaxies?} Monthly Notices of the Royal Astronomical Society, 487, 2278

\bibitem[{Z. Wang {et~al.}(2024)Wang, Shi, Yang, Li, Liu, \&
  Li}]{wang2024darkai}
Wang, Z., Shi, F., Yang, X., {et~al.} 2024, \bibinfo{title}{(DarkAI) Mapping
  the large-scale density field of dark matter using artificial intelligence,}
  Science China Physics, Mechanics \& Astronomy, 67, 219513

\bibitem[{R. Watkins {et~al.}(2009)Watkins, Feldman, \&
  Hudson}]{watkins2009consistently}
Watkins, R., Feldman, H.~A., \& Hudson, M.~J. 2009,
  \bibinfo{title}{Consistently large cosmic flows on scales of 100 h- 1 Mpc: a
  challenge for the standard $\Lambda$CDM cosmology,} Monthly Notices of the
  Royal Astronomical Society, 392, 743

\bibitem[{Q.-R. Yang {et~al.}(2025)Yang, Zhu, Yu, Mo, Zheng, \&
  Feng}]{yang2025width}
Yang, Q.-R., Zhu, W., Yu, G., {et~al.} 2025, \bibinfo{title}{On the width and
  profiles of cosmic filaments,} arXiv preprint arXiv:2507.02476

\bibitem[{D.~G. York {et~al.}(2000)York, Adelman, Anderson~Jr, Anderson, Annis,
  Bahcall, Bakken, Barkhouser, Bastian, Berman, {et~al.}}]{york2000sloan}
York, D.~G., Adelman, J., Anderson~Jr, J.~E., {et~al.} 2000,
  \bibinfo{title}{The sloan digital sky survey: Technical summary,} The
  Astronomical Journal, 120, 1579

\bibitem[{Y.~B. Zeldovich {et~al.}(1982)Zeldovich, Einasto, \&
  Shandarin}]{zeldovich1982giant}
Zeldovich, Y.~B., Einasto, J., \& Shandarin, S. 1982, \bibinfo{title}{Giant
  voids in the Universe,} Nature, 300, 407

\bibitem[{Y. Zhang {et~al.}(2024)Zhang, Guo, Yang, \&
  Wang}]{zhang2024statistical}
Zhang, Y., Guo, H., Yang, X., \& Wang, P. 2024, \bibinfo{title}{Statistical
  properties of filaments in the cosmic web,} Monthly Notices of the Royal
  Astronomical Society, 533, 1048

\end{thebibliography}
\bibliographystyle{aasjournalv7}

\end{document}